\documentclass[a4paper,prl,twocolumn,superscriptaddress,showkeys,floatfix]{revtex4}
\usepackage{graphicx}
\usepackage{amsmath}
\usepackage{amsthm}
\usepackage{amssymb}

\newcommand{\mean}[1]{\left\langle #1 \right\rangle}
\newcommand{\var}[1]{\mathrm{var}\!\left( #1 \right)}
\newcommand{\dd}{\mathrm{d}}
\newcommand{\erg}{\mathcal U}
\newcommand{\heat}{\mathcal C}
\DeclareMathOperator\atanh{atanh}

\begin{document}

\title{The discontinuity of the specific heat for the 5D Ising model}

\date{\today} 

\author{P. H. Lundow} 
\email{per.hakan.lundow@math.umu.se} 


\author{K. Markstr\"om}
\email{klas.markstrom@math.umu.se} 

\affiliation{ Department of mathematics and mathematical statistics,
  Ume\aa{} University, SE-901 87 Ume\aa, Sweden}

\begin{abstract}
  In this paper we investigate the behaviour of the specific heat
  around the critical point of the Ising model in dimension 5 to 7. We
  find a specific heat discontinuity, like that for the mean field
  Ising model, and provide estimates for the left and right hand
  limits of the specific heat at the critical point. We also estimate
  the singular exponents, describing how the specific heat approaches
  those limits.  Additionally, we make a smaller scale investigation
  of the same properties in dimension 6 and 7, and provide strongly
  improved estimates for the critical termperature $K_c$ in $d=5,6,7$
  which bring the best MC-estimate closer to those obtained by long
  high temperature series expanions.
\end{abstract}


\maketitle

\section{Introduction}
The Ising model in dimension $d=5$, being strictly larger than the
upper critical dimension $d_c=4$ of the model, has been studied by
many authors, and has been the focus of a long running debate
regarding its finite size scaling behaviour. We refer the reader to
\cite{Lundow2014249} for a discussion of that topic. However, even
regarding the infinite size limit of the model there are still
interesting open questions.  For $d>d_c$ it is rigorously known that
on the $d$-dimensional hypercubic lattice the critical exponents of
model takes their mean field values. This was first proven in
\cite{sokal:79, aizenman:81,aizenman:82,aizenman} and more recently
the newly developed rigorous lace expansion for the Ising model
\cite{sakai} made it possible to give a unified proof of these results
by a single method \cite{MR2430773}.  For the specific heat the
critical exponent $\alpha=0$ and the results of \cite{sokal:79} also
show a stronger result, namely that the specific heat is bounded at
the critical point.  However, the values of the critical exponents are
not the only properties which are characteristic for the phase
transition in the mean field version of the model, there is also a
discontinuity in the value of the specific heat at the critical point.
In fact, historically this discontinuity was once noted as one of the
first signs showing that mean field theory does not give a correct
description of phase transitions in dimension 2 and 3.  Hence it is
natural to ask if there is a similar discontinuity in the specific
heat for the Ising model above the upper critical dimension.

This question is also natural from another point of view. One way of
realizing the mean field version of the Ising model is to view it as
the thermodynamic limit of the Ising model on finite complete
graphs. The model on complete graphs has been studied rigorously in
great detail by mathematicians, both in the usual Ising form and the
equivalent Fortuin-Kateleyn random cluster representation,
\cite{MR1376340,luczak:06}.  Numerically it has been observed
\cite{5dart} that for $d=5$ the model on finite lattices with periodic
boundary conditions displays the same scaling behaviour inside the
critical scaling window as the Ising model on a complete graph.  Hence
it is also natural to ask if the behaviour of the thermodynamic limit
of the model on the complete graph and the hypercubic lattice will
also show the same type of behaviour at the critical point, in
particular if both have a discontinuous specific heat and how large
the jump at that discontinuity is.

In order to study this question we have done Monte Carlo simulation of
the Ising model on hypercubic lattices with periodic boundary
conditions, with the main effort for $d=5$ but with some data for
$d=6,7$ as well.  Using these data we first give improved estimates
for the critical temperatures in these dimensions. Next we find that
there is a jump in the specific heat and give estimates for the left
and right hand limits of the specific heat at the critical temperature
$K_c$. We also estimate the singular critical exponents, describing
how the specific heat approaches the limit values. As mentioned in
\cite{sokal:94} the singular exponents, unlike the critical ones, are
not expected to necessarily have the same value on the low and
high-temperature sides of the critical point, and we find that their
values are quite distinct. Finally we also note that as $d$ increases
the behaviour at the critical points seems to be approaching that of
the mean field limit, as expected.

The structure of the paper is as follows. After some definitions we
first give a derivation of the specific heat for the complete graphs,
and use it to give a description of the specific heat for the mean
field limit which is more detailed than the usual one.  Next we
present our numerical data for $d=5$, first for the critical
temperature $K_c$ and then for the specific heat near $K_c$. After
that we give a brief description of the corresponding results for
$d=6,7$, and finally we give some discussion of the observed results.

\section{Definitions and details}
For a given graph $G$ on $N$ vertices the Hamiltonian with
interactions of unit strength along the edges is
$\mathcal{H}=-\sum_{ij} S_i S_j$ where the sum is taken over the edges
$ij$. As usual the coupling $K=1/k_BT$ is the dimensionless inverse
temperature and we denote the thermal equilibrium mean by
$\mean{\cdots}$.

The we call the critical coupling $K_c$, and denote its normalised
form by $\varepsilon=(K-K_c)/K_c$ and the rescaled version $\kappa =
\sqrt{N} (K-K_c)/K_c$. As usual the magnetisation is $M=\sum_i S_i$
(summing over the vertices $i$) and the energy is $E=\sum_{ij}S_iS_j$
(summing over the edges $ij$). We let $m=M/N$, $U=E/N$ and
$\erg=\mean{U}$. The specific heat is defined as
\begin{equation}
  \heat=\frac{-\partial^2}{\partial T \partial K}\frac{\log
    Z}{N}=\frac{K^2}{N} \left(\mean{E^2}-\mean{E}^2\right)
\end{equation}

When the underlying graph is a $d$-dimensional grid graph of linear
order $L$ with periodic boundary conditions we mean it simply to be
the cartesian product of $d$ cycles on $L$ vertices, so that $N=L^d$.
When we refer to the complete graph we mean the graph where all pairs
of vertices are connected by an edge, thus having
$\binom{N}{2}=N(N-1)/2$ edges.  For $d=5$ we have collected data using
Wolff-cluster updating for $L=16$, $20$, $24$, $32$, $40$, $48$, $56$
and $64$. The number of measurements at each temperature near $K_c$
ranges from ca $15000$ for $L=64$ to more than $100000$ for
$L=16,20,24,32$. We will also re-use some extremely detailed data from
\cite{boundarypaper} for $L=6$, $8$, $10$ and $12$.

\section{The Ising model on the complete graph}
Recall that the complete graph $G_N$ on $N$ vertices is the graph with
$N$ vertices in which every pair of of distinct vertices is joined by
an edge.  The limit as $N\rightarrow \infty$ of the Ising model on
$G_N$ corresponds to the usual mean field Ising model. In order to be
able to make a detailed comparison with the $d$-dimensional Ising
model we will now derive an expression for the specific heat of the
mean field model in a neighbourhood of the critical point, instead of
the more common textbook version which only gives the jump exactly at
the critical point.

First note that it is an exercise to show that
$\mean{E}=(1/2)\mean{M^2}-N/2$ and, more importantly,
\begin{equation}\label{eq:varE}
  \var{E} = \frac{1}{4}\var{M^2} = \frac{1}{4}\left(\mean{M^4}-\mean{M^2}^2\right)
\end{equation}
This will come in handy when we compute
$\heat(K)=\lim_{N\to\infty}\heat(K,N)$ where
$\heat(K,N)=K^2\var{E}/N$.

It is shown in Ref.~\cite{pqpaper} that the magnetisation distribution
at coupling $K$ for a complete graph is
\begin{equation}
  \Pr(M=N-2k) = \frac{1}{\psi} q^{k(N-k)} \binom{N}{k}, \quad 0\le k \le N
\end{equation}
where $q=\exp(-2K)$. Since $\sum_k \Pr(M=N-2k) = 1$ this implicitly
defines $\psi$. When $q=N/(N+2)$ the distribution is precisely flat in
the middle, i.e. with $\Pr(M=-2)=\Pr(M=0)=\Pr(M=+2)$ (for even $N$)
and thus
\begin{equation}
  K_c=\frac{1}{N}-\frac{1}{N^2}+\frac{4}{3N^3} + \cdots
\end{equation}
constitutes an effective $K_c$. The appendix of Ref.~\cite{pqpaper}
provides detailed information on the moments of this magnetisation
distribution and we will apply this to get information on the energy
moments. We begin with the case of $q=N/(N+2) - 2\kappa/N^{3/2}$. With
$q=\exp(-2K)$ this corresponds to $\kappa = \sqrt{N}(K-K_c)/K_c +
O(1/N)$, i.e. we move around inside the scaling window with the
temperature parameter $\kappa$. Using Lemma A6 of Ref.~\cite{pqpaper}
(after setting $a=-2\kappa$) we can now easily obtain the asymptotic
form of the $\ell$th moment as
\begin{equation}\label{eq:mom}
  \mean{|M|^{\ell}} \sim 
  \frac{N^{3\ell/4} 2^{\ell}
      \int\limits_{-\infty}^{\infty}|x|^{\ell}R(\kappa,x)\dd x
    }{
      \int\limits_{-\infty}^{\infty}R(\kappa,x)\dd x
    }
\end{equation}
where $R(\kappa,x)=\exp(2\kappa x^2-4x^4/3)$.  Plugging this into
Equation~\eqref{eq:varE} and evaluating the integrals we can obtain a
formula for $\heat$.  However, in the special case $\kappa=0$, we get
the very simple
\begin{equation}
  \heat=\frac{3}{4}-\frac{6\pi^2}{\Gamma(1/4)^4}\approx 0.4072901
\end{equation}
The local maximum of $\heat$ can now be computed numerically to lie at
$\kappa^*=2.2568473919660\ldots$ and the value at this point is
$\heat_{\max}=1.6572974585496\ldots$. We note also the limits
$\lim_{\kappa\to\infty}\heat(\kappa)=3/2$ and
$\lim_{\kappa\to-\infty}\heat(\kappa)=0$.

To continue with the case outside the scaling window we set
$q=(n-2\varepsilon)/(n+2)$ which, since $q=\exp(-2K)$, gives us
$\varepsilon=(K-K_c)/K_c + O(1/N)$, our normalised temperature.  In
the high-temperature case, i.e. for $\varepsilon<0$, we get from Lemma
A9 (setting $a=-2\varepsilon$) of Ref.~\cite{pqpaper} that
\begin{equation}\label{eq:mom1}
  \mean{|M|^{\ell}} = \frac{
    N^{\ell/2} 2^{\ell}
    \int\limits_{-\infty}^{\infty} |x|^{\ell}\exp(2\varepsilon x^2)\dd x
  }{
    \int\limits_{-\infty}^{\infty}\exp(2\varepsilon x^2)\dd x
  }
\end{equation}
so that $\heat(\varepsilon,N)=O(1/N)$ and thus
$\heat(\varepsilon,\infty)=0$. This could be interpreted as
$\theta^-=1$. The case $\varepsilon=0$ was treated above as
$\kappa=0$.

The low-temperature case $\varepsilon>0$ is a little more tricky. Let
$\mu=\mean{|m|}$, where $0<\mu<1$, denote the normalised (spontaneous)
magnetisation and note that the magnetisation distribution has a peak
at $M_{\mathrm{peak}}=\pm\mu N$ having width $O(\sqrt{N})$.  Moving
$x\sqrt{N}$ magnetisation steps away from $M_{\mathrm{peak}}$ we are
at the new magnetisation $M_x$ where
$|M_x|=2\sqrt{N}|x+\mu\sqrt{N}/2|$. Lemma A13 says that the ratio
$\Pr(M_x)/\Pr(M_{\mathrm{peak}})$ is asymptotically
\begin{equation}
  R(\mu,x) = \exp\left\{2x^2\left(\frac{1}{\mu^2-1} + \frac{\atanh(\mu)}{\mu}\right)\right\}
\end{equation}
and the $\ell$th moment then becomes
\begin{equation}
  \mean{|M|^{\ell}} \sim \frac{
    N^{\ell/2}
    2^{\ell}\int\limits_{-\infty}^{\infty}
    \left|x+\frac{\mu\sqrt{N}}{2}\right|^{\ell}R(\mu,x)\dd x
  }{
    \int\limits_{-\infty}^{\infty} R(\mu,x)\dd x
  }
\end{equation}

\begin{figure}
  \begin{center}
    \includegraphics[width=0.483\textwidth]{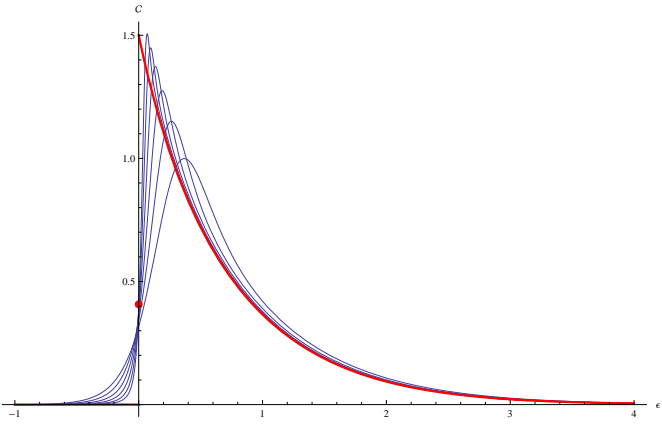}
  \end{center}
  \caption{(Colour on-line) $\heat(\varepsilon,N)$ versus
    $\varepsilon$, with $\varepsilon=(K-K_c)/K_c$ for finite $N=32$,
    $64$, $128$ $256$, $512$, $1024$ (increasing blue curves) and
    the complete graph limit case $N\to\infty$ (red thick curve). The
    red dot is the limit at $\varepsilon=0$.}\label{fig:Ckn}
\end{figure}

Using Equation~\eqref{eq:varE} the specific heat limit, expressed in
$\mu$, collapses into the simple form
\begin{equation}\label{eq:heat}
  \heat(\mu) = \frac{\mu^3 - \mu^5}{\mu + (\mu^2-1)\atanh(\mu)}
\end{equation}
Next, Lemma A11~\cite{pqpaper}, after setting $a=-2\varepsilon$, says
that $\mu$ depends on $\varepsilon$ as asymptotically
\begin{equation}\label{eq:mu}
  \varepsilon = \frac{\atanh(\mu)}{\mu} - 1
\end{equation}
In combination with Equation~\eqref{eq:heat} this defines implicitly
the limit specific heat in terms of $\varepsilon$. Taking the
composition of the series expansion of Equation~\eqref{eq:heat} and
the inverse series expansion of Equation~\eqref{eq:mu} we obtain at
last
\begin{equation}\label{eq:series}
  \heat(\varepsilon) \sim \frac{3}{2} - \frac{12}{5}\varepsilon + 
    \frac{438}{175}\varepsilon^2 - \frac{432}{175}\varepsilon^3 
    + \frac{166104}{67375}\varepsilon^4 + \cdots
\end{equation}
Plotting the numerical evaluation of \eqref{eq:mu} and \eqref{eq:heat}
we get the Figure~\ref{fig:Ckn} where the limit and some finite cases
are shown. 

\section{The 5-dimensional case}
We now come to our Monte Carlo results for $d=5$.  As noted in
\cite{boundarypaper} some of the indicators used by other authors to
study the critical behaviour for $d=5$ are very sensitive to the exact
value of the critical temperature $K_c$.  With this in mind we will
first present a new way of obtaining highly precise estimates for
$K_c$ and use it to derive the estimate which we will use in our later
analysis.

\subsection{An improved method for estimating $K_c$}
Our improved estimate of $K_c$, suitable for $d\ge 4$, is based on a
careful study of the magnetisation distribution, i.e. $\Pr(M)$. The
approach is simple but assumes that all measurements of $M$ were
stored for each temperature during the sampling process. After
normalising these values as $x=M/N^{3/4}$ we put them in bins of
reasonable width, in our case $0.20$, thus giving us a histogram.
There are of course several different binning methods to choose from,
but for simplicity we have chosen to use a fixed bin width which is
roughly what the Freedman-Diaconis method (twice the interquartile
range divided by the cube root of the number of measurements)
prescribes when the distribution is near $K_c(L)$ (see below) for the
weakest data set (i.e. for $L=64$).

The simple distribution density function $f(x)=\phi_0\,\exp(\phi_2
x^2+\phi_4 x^4)$ is then fitted to this histogram.  Since $\phi_2$ for
all intents and purposes depends linearly on $K$ inside the scaling
window (see Fig.~3 of \cite{pqpaper2}), we fit a straight line to the
data points (at least seven) on the interval corresponding to
$-0.7<\phi_2<0.7$ and solve $\phi_2(K)=0$. This point constitutes an
effective critical temperature $K_c(L)$ scaling as $K_c(L)-K_c\propto
L^{2-d}$, see~\cite{brezin:85}.

Ideally the density function $f(x)$ should also contain a correction
factor $(1+\lambda_2 x^2 + \lambda_4 x^4+\lambda_6 x^6+\ldots)$ but
the coefficients $\lambda_i$ will vanish with increasing $L$. For
$L\ge 16$, especially near $K_c(L)$, they will not contribute
significantly to $f(x)$ and can in any case not be discerned with the
data we rely on here.  See~\cite{pqpaper,pqpaper2} for a considerably
more detailed study of the scaling behaviour of the magnetisation
distribution.

\begin{figure}
  \begin{center}
    \includegraphics[width=0.483\textwidth]{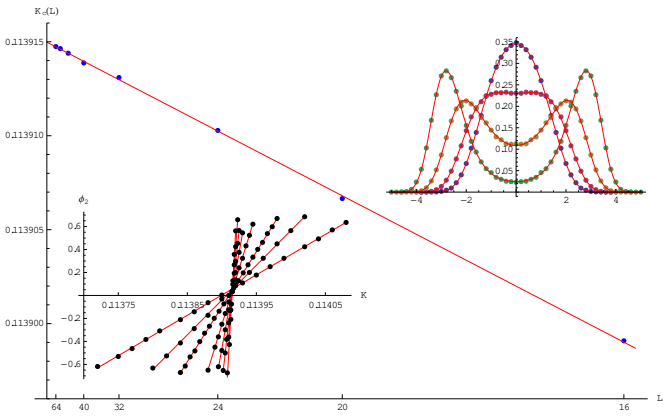}
  \end{center}
  \caption{(Colour on-line) $K_c(L)$ versus $1/L^3$ for $L=16$, $20$,
    $24$, $32$, $40$, $48$, $56$ and $64$. The fitted line is
    $0.11391498-0.0654x$. The lower inset shows the fitted parameter
    $\phi_2$ versus $K$ for the same $L$. The upper inset shows the
    normalised magnetisation distribution $\Pr(M/N^{3/4})$ for
    $L=32$ at $K=0.1139$, $0.113915$, $0.11393$ and $0.113945$
    resulting in $\phi_2=-0.251$, $0.036$, $0.323$ and $0.621$
    respectively of the fitted $f(x)$ (red curves).}\label{fig:KcL}
\end{figure}

In Figure~\ref{fig:KcL} we show $K_c(L)$ versus $1/L^3$ together with
an inset showing the $M/N^{3/4}$-distribution for $L=32$ at different
values of $K$ and another inset showing how $\phi_2$ depends on $K$
for the different system sizes.  A line fit gives that
$K_c(L)=0.11391498(2)-0.0654(2)\,L^{-3}$.  The coefficients and their
error estimates are here based on the median and interquartile range
of the fitted coefficients when deleting each data point in turn from
the line fit. The estimate $K_c=0.11391498(2)$ is within the error
bars of earlier estimates~\cite{LBB:99,BL:97} but adds another digit
to the accuracy.  This technique for estimating $K_c$ is quite
robust to variations in the various parameters. For example, changing
the distribution bin widths to $0.15$ or using $\phi_2$ data for
$-0.6<\phi_2<0.6$ keeps the resulting $K_c$ within the stated error
bars.

\subsection{Specific heat discontinuity}
\begin{figure}
  \begin{center}
    \includegraphics[width=0.483\textwidth]{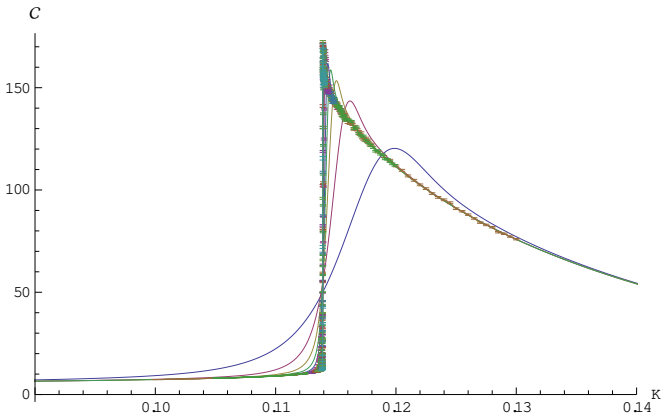}
  \end{center}
  \caption{(Colour on-line) $\heat(K,L)$ for $L=4$, $6$, $8$, $10$,
    $12$, $16$, $20$, $24$, $32$, $40$, $48$, $56$ and
    $64$.}\label{fig:Cwide}
\end{figure}

Consider Figure~\ref{fig:Cwide} where we plot the specific heat for
$4\le L\le 64$ for a wide temperature range. Clearly there is an
envelope curve containing the limit specific heat.  From the
individual $\heat(K,L)$ functions we extract the limit function
$\heat(K,\infty)=\lim_{L\to\infty}\heat(K,L)$ from points where the
function for increasing $L$ agree.  We thus assume that there is a
$K_{\min}(L)$ such that if $L'\ge L$ and $K>K_{\min}(L)>K_c$ then
$\heat(K,L')=\heat(K,\infty)$. Analogously we assume there is a
$K_{\max}(L)$ such that $\heat(K,L')=\heat(K,\infty)$ when
$0<K<K_{\max}(L)<K_c$ and $L'\ge L$.
As an example, taking for example
$L_1=16$ and $L_2=20$ we note that $\heat(K,16)=\heat(K,20)$ when
$K>K_{\min}=0.1148$, where $K_{\min}$ of course depends on the chosen
$L_1$ and $L_2$. On the high-temperature side of $K_c$ we find that
$\heat(K,16)=\heat(K,20)$ when $K<K_{\max}(16)=0.1120$. Thus we treat
the measured data for $\heat(K,L)$ as the asymptotic $\heat(K,\infty)$
when $L\ge 16$ and $K<0.1120$ or $K>0.1148$. An increasing sequence of
pairs of $L_1,L_2$ gives a sequence of $K_{\min}$ and $K_{\max}$ that
both approach $K_c$.  The individual $K_{\min}$ and $K_{\max}$ were
found by simply comparing pairwise plots of $\heat(K,L)$.

\begin{figure}
  \begin{center}
    \includegraphics[width=0.483\textwidth]{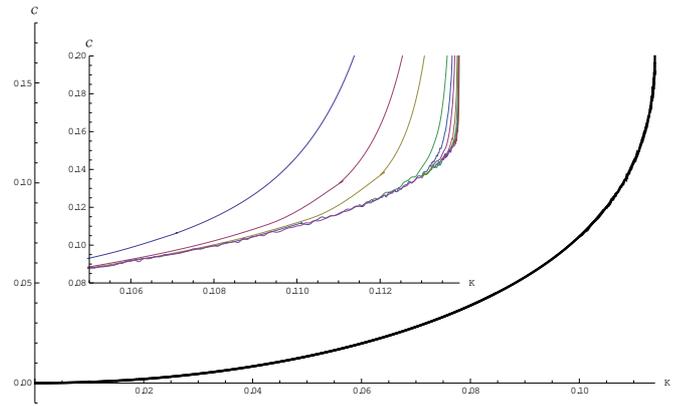}
  \end{center}
  \caption{(Colour on-line) $\heat(K,L)$ versus $K$ for
    $K<K_{\max}(L)$ for $L=6$, $8$, $10$, $16$, $20$, $24$, $32$,
    $40$, $48$, $56$ and $64$. The curve consists of more than 500
    points and error bars are not shown. The inset shows a zoomed-in
    version with 3rd order interpolations through all data
    points. Data for increasing $L$ start to deviate from envelope
    curve as we move closer to $K_c$.}\label{fig:asyhi}
\end{figure}

\begin{figure}
  \begin{center}
    \includegraphics[width=0.483\textwidth]{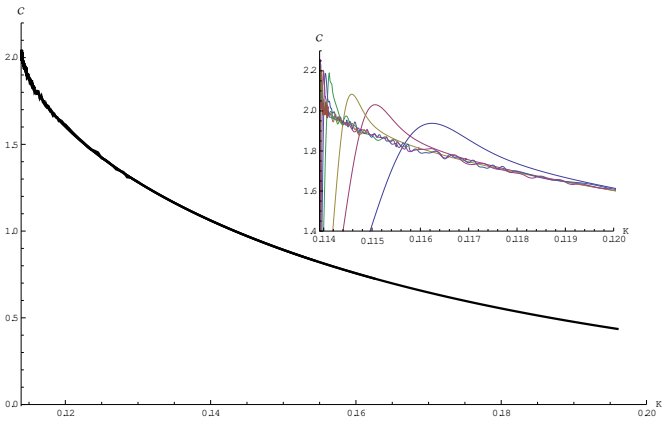}
  \end{center}
  \caption{(Colour on-line) $\heat(K,L)$ versus $K$ for
    $K>K_{\min}(L)$ for $L=6$, $8$, $10$, $16$, $20$, $24$, $32$,
    $40$, $48$, $56$ and $64$.The curve consists of more than 500
    points and error bars are not shown. The inset shows a zoomed-in
    version of 3rd order interpolations through all data points. Data
    for increasing $L$ deviate from envelope curve as we move closer
    to $K_c$.}\label{fig:asylo}
\end{figure}

In Figure~\ref{fig:asyhi} we show the individual $\heat(K,L)$ for
$K<K_{\max}(L)$ pieced together into one plot for a range of $L$ and
Figure~\ref{fig:asylo} shows the coresponding data for
$K>K_{\min}(L)$. Their insets shows the data without removing the
finite size behaviour and clearly demonstrate the presence of a limit
enveloping curve. Having removed any finite size effects, such as the
local maximum for each $L$, the remaining points are in effect
estimates of the asymptotic $\heat(K,\infty)$ for
$0<K<K_{\max}(64)=0.11388$ and $K>K_{\min}(64)=0.11394$.

In order to estimate the left- and right-limit we take the
limit curves and fit a simple expression of the form
\begin{equation}\label{eq:fit}
  A_0 + A_1 x^{\theta} (1 + B_1 x + B_2 x^2 + \ldots)
\end{equation}
to $\heat(K,\infty)$ where $x=|K-K_c|/K_c$. We will use
$\pm$-superscripts to denote the left- and right-limit as $x\to
0$. This provides left- and right-limits ($A_0^-$ and $A_0^+$), the
dominating correction term exponents ($\theta^-$ and $\theta^+$) and a
sequence of correction terms. Following \cite{sokal:94} we call
$\theta^-$ and $\theta^+$ the \emph{singular exponents} of the
specific heat, since they describe the behaviour of the singular part
of the specific heat.  We are not aware of any prescribed form of the
correction terms from earlier studies so these will simply be the
effective terms.

Using Mathematica's built-in FindFit-function we fit the
high-temperature limit curve to \eqref{eq:fit} using both one, two and
three correction terms and find excellent agreement in the resulting
values of the singular exponent $\theta^-$, measuring
$\theta^-=0.40(1)$. The first two coefficients $A_0^-$ and $A_1^-$
also strongly agree when adding more correction terms.  However,
having first established a strong candidate exponent we now simply fix
this to $\theta^-=0.4$ and use \eqref{eq:fit}, again trying one, two
and three correction terms.  Based on this we find an effective fit
\begin{equation}\label{eq:hi}
  \heat^-(x)= 0.1697(2) - 0.231(1) x^{0.40} (1 - 0.26(1) x)
\end{equation}
where $x=-\varepsilon=(K_c-K)/K$ and $0<K<K_c$.  Adding more
correction terms does not improve the fit.  The error bars reflect how
the coefficients change when adding one or two more terms.

Repeating this exercise for the low-temperature side the
FindFit-function suggests $\theta^+=0.60(2)$ and again the leading
coefficients agree using one, two and three correction terms.  Setting
$\theta^+=0.60$ gives us the effective fit
\begin{equation}\label{eq:lo}
  \heat^+(x) = 2.040(1) - 2.58(1) x^{0.60} (1 - 0.36(2) x)
\end{equation}
where $x=\varepsilon=(K-K_c)/K$ and $K>K_c$. As before, the error bars
reflect how the coefficients change when adding correction terms.  We
now put \eqref{eq:hi} and \eqref{eq:lo} to the test by taking log-log
plots of the measured $\heat(K,\infty)$ when subtracting the
respective limit $A_0^{\pm}$.

\begin{figure}
  \begin{center}
    \includegraphics[width=0.483\textwidth]{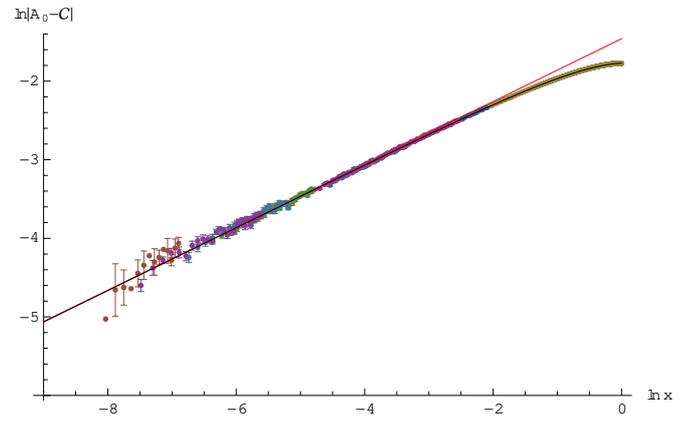}
  \end{center}
  \caption{(Colour on-line) $\log(0.1697-\heat(K,\infty))$ versus
    $\log(x)$, with $x=(K_c-K)/K_c$, for $L=6$, $8$, $10$, $16$, $20$,
    $24$, $32$, $40$, $48$, $56$ and $64$ (larger $L$ to the left in
    the figure) together with the fitted $\heat^-(x)$ (black curve,
    hard to see) of Eq.~\eqref{eq:hi} and the asymptote
    $0.1697-0.231x^{0.4}$ (red line with slope
    $0.40$).}\label{fig:logloghi}
\end{figure}

\begin{figure}
  \begin{center}
    \includegraphics[width=0.483\textwidth]{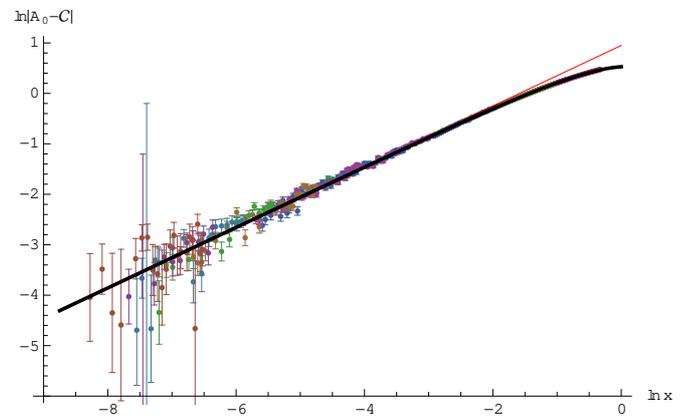}
  \end{center}
  \caption{(Colour on-line) $\log(2.04-\heat(K,\infty))$ versus
    $\log(x)$, with $x=(K-K_c)/K_c$, for $L=6$, $8$, $10$, $16$, $20$,
    $24$, $32$, $40$, $48$, $56$ and $64$ (larger $L$ to the left in
    the figure) together with the fitted $\heat^+(x)$ (black curve,
    hard to see) of Eq.~\eqref{eq:lo} and the asymptote
    $2.04-2.58x^{0.60}$ (red line with slope $0.60$.}\label{fig:logloglo}
\end{figure}

Beginning with the high-temperature case, in Figure~\ref{fig:logloghi}
we show $\log(0.1697-\heat(K,\infty))$ versus $\log(x)$, where
$x=(K_c-K)/K_c$, together with $\heat^-(x)$ of \eqref{eq:hi} (black
curve) and the asymptote $0.1697-0.231 x^{0.40}$ (red line). The rather
small error bars suggest a good quality of the fit.  Analogously, on
the low-temperature side, we show in Figure~\ref{fig:logloglo}
$\log(2.04-\heat(K,\infty))$ versus $\log(x)$, where $x=(K-K_c)/K_c$,
together with $\heat^+(x)$ of \eqref{eq:lo} (black curve) and the
asymptote $2.04-2.58 x^{0.60}$ (red line). For $L=56, 64$ the error
bars are now quite pronounced.  For the smaller $L$ the error bars are
considerably more benign.

\begin{figure}
  \begin{center}
    \includegraphics[width=0.483\textwidth]{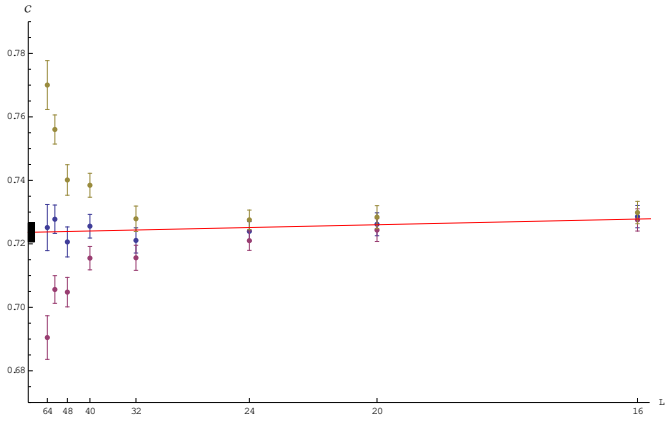}
  \end{center}
  \caption{(Colour on-line) $\heat(K,L)$ versus $1/L^{5/2}$, for $L=16$,
    $20$, $24$, $32$, $40$, $48$, $56$ and $64$ together with the
    fitted red line $0.724+4.19 x$, where $x=1/L^{5/2}$.  The points
    trending upwards are for $K=0.1139152$, the points trending
    downwards are for $K=0.1139148$ and the middle set of points are
    for $K_c=0.11391498$. The rectangle on the $y$-axis indicates the
    estimate $\heat(K_c,\infty)=0.724(3)$. }\label{fig:Ccrit}
\end{figure}

Finally we estimate the value of $\heat(K_c,\infty)$, which, of
course, does not have to coincide with any of the $A_0^{\pm}$.  As it
turns out this value is quite distinct from both limits.  In
Figure~\ref{fig:Ccrit} we show a zoomed-in plot of $\heat(K,L)$ over a
range of $L$ for three fixed $K$-values, $K=0.11391498$ (i.e. the
estimated $K_c$), $K=0.1139148$ and $K=0.1139152$. The $\heat(K,L)$ were
found by interpolating the data points. As the plot 
demonstrates, there is a clear upwards trend in the values for
$K=0.1139152$ and a clear downwards trend for $K=0.1139148$, whereas
the middle value shows no clear trend.  A fitted line on the points
for $L\ge 16$ suggests $\heat(K_c,\infty)=0.724(3)$.  The error bar of
this value is obtained by allowing the value of $K$ to vary inside the
error bar of $K_c$ (2 steps in the 8th digit) and repeat the line fit
to the new points.

\begin{figure}
  \begin{center}
    \includegraphics[width=0.483\textwidth]{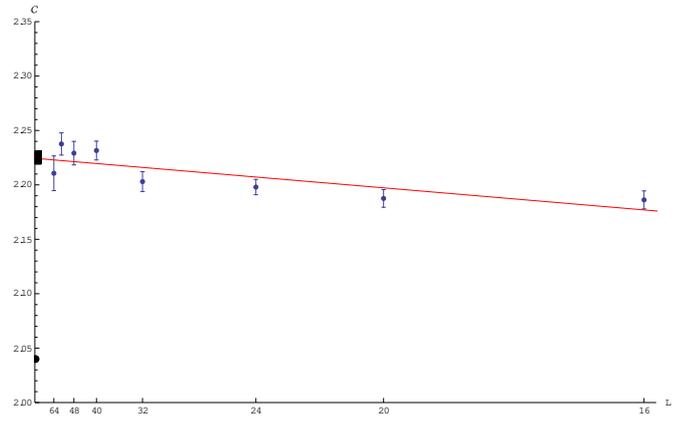}
  \end{center}
  \caption{(Colour on-line) $\max_K \heat(K,L)$ versus $1/L^{5/2}$,
    for $L=16$, $20$, $24$, $32$, $40$, $48$, $56$ and $64$ together
    with the fitted red curve $2.225-48.6 x$, where $x=1/L^{5/2}$. The
    rectangle on the $y$-axis indicates the limit estimate
    $\heat_{\max}=2.225(6)$. The point at $y=2.04$ is the right-limit
    $\heat^+(0)=2.04$ of Eq.~\eqref{eq:lo}.}\label{fig:Cmax}
\end{figure}

The local maximum of $\heat(K,L)$, see Figure~\ref{fig:asylo}, also
takes its own limit value, i.e. $\heat_{\max}=\lim_{L\to\infty} \max_K
\heat(K,L)$ does not coincide with the right-hand limit $\heat^-$.  In
Figure~\ref{fig:Cmax} we show the estimated maximum for each $L$ and
the right-hand limit $2.04$ found above \eqref{eq:lo}. It appears very
unlikely that they should coincide for large $L$. The fitted line,
based on $L\ge 16$, suggests a limit $\heat_{\max}=2.225(6)$ where the
error bar is based on the variability of the constant term of fitted
lines (with $x=1/L^{5/2}$) with one point removed from the data set
$L\ge 16$. We estimate that the maximum is located at $K_{\max} = K_c
+ 1.860(2)K_c/L^{5/2}$, i.e., at $\kappa=1.860(3)$, with the error bar
obtained as before by removing individual points for $L\ge 16$ when
fitting a line through the origin (since we know $K_{\max}\to K_c$).
We will plot $\heat$ versus $\kappa$ later.

\begin{figure}
  \begin{center}
    \includegraphics[width=0.483\textwidth]{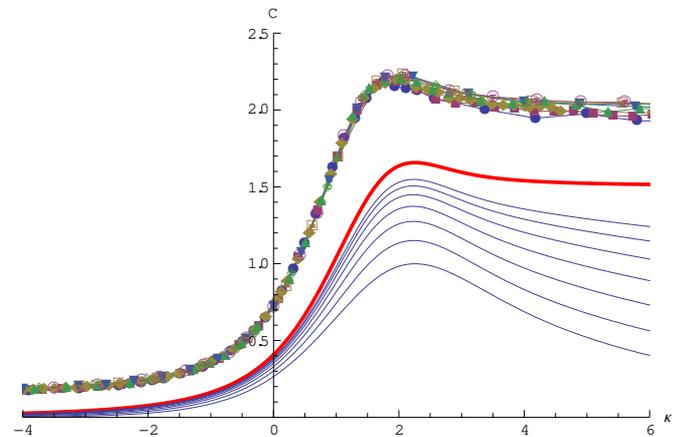}
  \end{center}
  \caption{(Colour on-line) $\heat$ versus $\kappa$, with
    $\kappa=\sqrt{N}(K-K_c)/K_c$ for finite $N=32$, $64$, $128$ $256$,
    $512$, $1024$, $2048$ (increasing blue curves) and the complete
    graph limit case $N\to\infty$ (red thick curve). The upper set of
    points shows the same for 5D sampled data points for $L=16$, $20$,
    $24$, $32$, $40$, $48$, $56$ and $64$. Error bars are shown but of
    the same size as the points.}\label{fig:Ckappa}
\end{figure}

We can now make a comparison of the behaviour of the 5-dimensional
model and that of the mean field case, as derived in the previous
section.  We first consider the scaling window, in
Figure~\ref{fig:Ckappa} we show a plot of $\heat(\kappa)$ for a range
of $N$ and the limit case, together with our data for $d=5$.  As we
can see that maximum specific heat for the mean field limit is lower
than the values for $d=5$, but the general shape of the curves are
nonetheless quite similar.

\begin{figure}
  \begin{center}
    \includegraphics[width=0.483\textwidth]{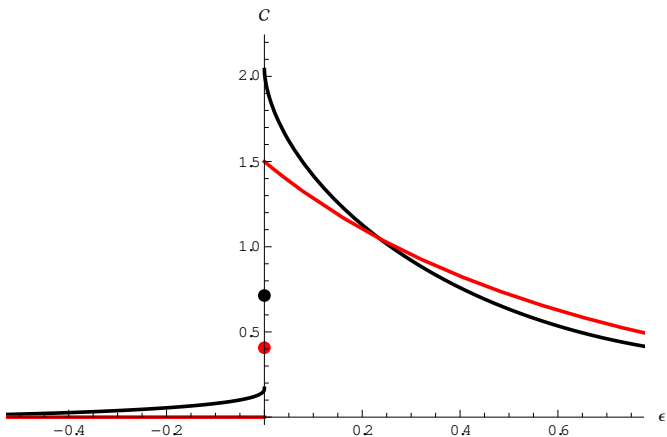}
  \end{center}
  \caption{(Colour on-line) The limit specific heat
    $\heat(\varepsilon$ with $\varepsilon=(K-K_c)/K_c$ for the
    complete graph (red, $y=1.5$, $y=0.407$ and $y=0$ at $y$-axis) and
    the 5D case (black, $y=2.04$, $y=0.72$ and $y=0.17$ at
    $y$-axis).}\label{fig:Cboth}
\end{figure}

Next we look at the thermodynamic limit. In Figure~\ref{fig:Cboth} we
show the specific heat limit for both the complete graph and $d=5$ in
the same plot.  As we just noted, the value for the mean field are
lower than those for $d=5$ when we are sufficiently close to
$\epsilon=0$. We can also see the difference in the singular exponents
between the two models, with the mean field case approaching the line
$\epsilon=0$ at an angle and the $d=5$ case instead approaching it
tangentially.

\section{The model in dimensions 6 and 7}
As the dimension $d$ increases we should see the specific heat
approach that of the complete graph. We also collected data for $d=6$
and $d=7$ and tried to estimate the singular exponents $\theta^+$ and
$\theta^-$. However, these data rely on considerably smaller systems;
$L=4,6,8,10,12,16,20$ for $d=6$ and only $L=4,6,8,10$ for $d=7$.  The
process is the same as we used above for $d=5$ and we will simply
state the resulting estimates of the various parameters.

\begin{figure}
  \begin{center}
    \includegraphics[width=0.483\textwidth]{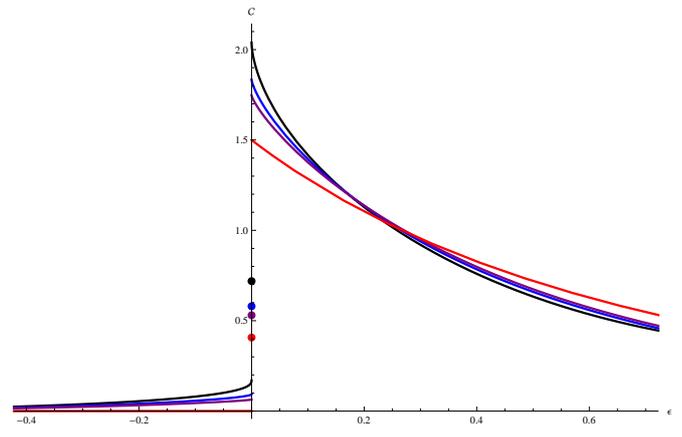}
  \end{center}
  \caption{(Colour on-line) The limit specific heat
    $\heat(\varepsilon)$ with $\varepsilon=(K-K_c)/K_c$ for $d=5$
    (black), $d=6$ (blue) and $d=7$ (purple) and the complete graph
    case (red), trending downwards at the $y$-axis, as do the points
    at $\varepsilon=0$.}\label{fig:Call}
\end{figure}

\begin{figure}
  \begin{center}
    \includegraphics[width=0.483\textwidth]{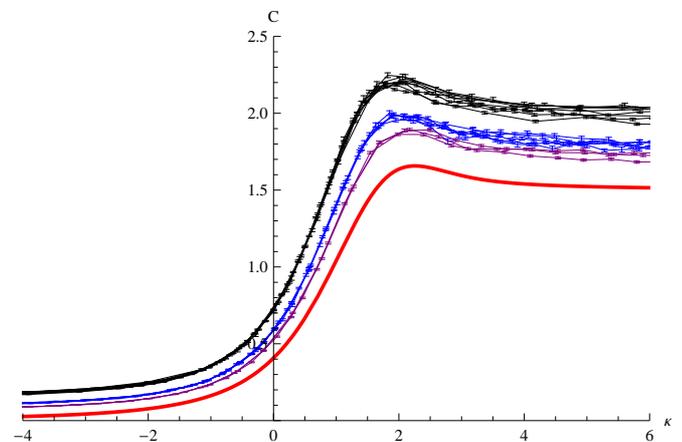}
  \end{center}
  \caption{(Colour on-line) The specific heat $\heat(\kappa)$ with
    $\kappa=\sqrt{N}(K-K_c)/K_c$. Trending downwards we see $d=5$
    (black, $L\ge 16$), $d=6$ (blue, $L\ge 8$) and $d=7$ (purple,
    $L\ge 6$) and the complete graph limit
    (red).}\label{fig:Ckappa-all}
\end{figure}

For $d=6$ we estimate $K_c=0.0922982(3)$ and
$\heat(K_c,\infty)=0.58(1)$.  This value of $K_c$ deviates somewhat
from the older Monte Carlo estimates, as surveyed in \cite{berche:08},
which have tended to be close to 0.09229, but agrees well with the
more recent estimate $0.092298(1)$ \cite{PhysRevE86011139}, coming
from the longest series expansion results to date.  The limit specific
heat is
\begin{equation}
  \heat = 
  \begin{cases}
    1.833 - 2.61 \varepsilon^{0.75} (1-0.57 \varepsilon) & \varepsilon > 0 \\
    0.58 & \varepsilon=0\\
    0.0927 - 0.148(-\varepsilon)^{0.60} (1+0.37\varepsilon) & \varepsilon < 0
  \end{cases}
\end{equation}
For $d=7$ we estimate $K_c=0.0777086(8)$, this is again closer to the
series based estimate from \cite{PhysRevE86011139} than the
MC-estimates from \cite{berche:08}, and
$\heat(K_c,\infty)=0.53(2)$. The limit specific heat is
\begin{equation}
  \heat = 
  \begin{cases}
    1.75 - 2.47 \varepsilon^{0.80} (1-0.56 \varepsilon+0.14\varepsilon^2) & \varepsilon > 0 \\
    0.53 & \varepsilon=0\\
    0.064 - 0.12(-\varepsilon)^{0.75} (1+0.45\varepsilon) & \varepsilon < 0
  \end{cases}
\end{equation}
In both cases the uncertainty in the coefficients is in the last
stated digit.  Combining $d=5,6,7$ and the complete graph case we plot
them all in Figure~\ref{fig:Call}. Inside the scaling window, that is,
with respect to $\kappa=\sqrt{N}(K-K_c)/K_c$, we can also clearly see
how the specific heat for finite-dimensional systems approach the
complete graph limit case. In Figure~\ref{fig:Ckappa-all} we plot
$\heat(\kappa,L)$ for several linear orders $L$ for $d=5,6,7$ and the
complete graph.

\section{Discussion}
We have estimated the critical behaviour of the specific heat of the
5-dimensional case and derived the limit curve for the complete
graph. The singular exponents for $d=5$ were found to differ for the
high- and low-temperature case. To summarise, for
$\varepsilon=(K-K_c)/K_c$ we estimate in the 5d case that for
$L\to\infty$ the specific heat behaves as
\begin{equation}
  \heat = 
  \begin{cases}
    2.040 - 2.58 \varepsilon^{0.60} (1-0.36 \varepsilon) & \varepsilon > 0 \\
    0.724 & \varepsilon=0\\
    0.1697 - 0.231(-\varepsilon)^{0.40} (1+0.26\varepsilon) & \varepsilon < 0
  \end{cases}
\end{equation}
The singular exponents are thus $\theta^+=0.60$ and $\theta^-=0.40$
for $d=5$.  As $d\to\infty$ we expect these exponents to approach
those of the complete graph where we find $\theta^+=\theta^-=1$.  The
exact series expansion of the limit specific heat for the complete
graph is
\begin{equation}
  \heat = 
  \begin{cases}
    \frac{3}{2} - \frac{12}{5} \varepsilon (1- \frac{73}{70}\varepsilon 
    + \frac{36}{35}\varepsilon^2+\cdots) & \varepsilon > 0 \\ 
    0.40729006421665228\ldots & \varepsilon=0 \\
    0 & \varepsilon < 0
  \end{cases}
\end{equation}

An open question which would be interesting to settle is how the left
and right hand limits of the specific heat for the $d$-dimensional
Ising model scales. We expect the limits to approach those of the mean
field model as $d\rightarrow\infty$ but we do not yet know how it
approaches those values.  That the $d\rightarrow\infty$ limit of the
value exactly at $K_c$ should be the same as the mean field value is
far from obvious and would also be worth further investigation.
Similarly we would like to know the scaling with $d$ of the left and
right singular exponents. We expect both of them to approach 1, but in
which way?

\section{Acknowledgements}
The simulations were performed on resources provided by the Swedish
National Infrastructure for Computing (SNIC) at High Performance
Computing Center North (HPC2N) and at Chalmers Centre for
Computational Science and Engineering (C3SE).


\end{document}